\documentclass[superscriptaddress,aps,pra,twocolumn,amssymb,amsfonts,showpacs]{revtex4}
\usepackage{times}
\usepackage{amsmath}

\usepackage{color}
\usepackage{graphics,graphicx}

\newcommand{\shortqph}[1]{}

\providecommand{\ignore}[1]{}

\newcommand{\mCo}[1]{\textcolor{blue}{}}
\newcommand{\hC}[1]{\textcolor{red}{}}

\def\openone{\leavevmode\hbox{\small1\kern-3.8pt\normalsize1}}
\def\RR{{\rm I\kern-.2emR}}

\def\openone{\leavevmode\hbox{\small1\kern-3.8pt\normalsize1}}
\def\RR{{\rm I\kern-.2emR}}

\def\cn{{\cal N}}
\def\ch{{\cal H}}
\def\cs{{\cal S}}

\def\cp{{\cal P}}

\def\cO{{\cal O}}

\def\cz{{\cal Z}}
\def\co{{\cal O}}

\providecommand{\ignore}[1]{}

\newcommand{\ket}[1]{| #1 \rangle}
\newcommand{\bra}[1]{\langle #1 |}

\newcommand{\bitem}{\begin{itemize}}
\newcommand{\eitem}{\end{itemize}}
\newcommand{\benum}{\begin{enumerate}}
\newcommand{\eenum}{\end{enumerate}}
\newcommand{\beq}{\begin{equation}}
\newcommand{\eeq}{\end{equation}}
\newcommand{\beqa}{\begin{eqnarray}}
\newcommand{\eeqa}{\end{eqnarray}}
\newtheorem{definition}{Definition}

\newtheorem{proposition}{Proposition}

\newcommand{\bproof}{\begin{proof}}
\newcommand{\eproof}{\end{proof}}
\newcommand{\bprop}{\begin{proposition}}

\newcommand{\bdef}{\begin{definition}}

\def\one{{\mathchoice {1\mskip-4mu {\text {\rm l}}} {
1\mskip-4mu {\text {\rm l}}} {1\mskip-4.5mu {\text {\rm l}}} {1\mskip-5mu
{\text {\rm l}}}}}

\def\({\left(}
\def\){\right)}

\begin{document}

\title{Quantum Simulated Annealing}

\author{R. D. Somma}
\email{somma@lanl.gov}

\affiliation{Perimeter Institute for Theoretical Physics, Waterloo, ON N2L 2Y5, Canada }

\author{S. Boixo} \affiliation{Los Alamos National Laboratory, Los
  Alamos, NM 87545, USA} \affiliation{Department of Physics and
  Astronomy, University of New Mexico, Albuquerque, NM 87131, USA}
\author{H. Barnum} \affiliation{Los Alamos National Laboratory, Los
  Alamos, NM 87545, USA}

\date{\today}
\begin{abstract}
  We develop a quantum algorithm to solve combinatorial optimization
  problems through quantum simulation of a classical annealing
  process.  Our algorithm combines techniques from quantum walks,
  quantum phase estimation, and quantum Zeno effect. It can be viewed
  as a quantum analogue of the discrete-time Markov chain Monte Carlo
  implementation of classical simulated annealing.  Our implementation
  requires order $1/\sqrt{\delta}$ operations to find an optimal
  solution with bounded error probability, where $\delta$ is the
  minimum spectral gap of the stochastic matrix used in the classical
  simulation. The quantum algorithm outperforms the classical one,
  which requires order $1/\delta$ operations.
\end{abstract}


\maketitle


\section{Introduction}
Combinatorial optimization problems (COPs) such as the traveling
salesman problem are important in almost every branch of science, from
computer science to statistical physics and computational
biology~\cite{CCP98}.  A COP consists of a family of {\em instances}
of the problem; each instance is an optimization problem, to minimize
(or maximize) some objective function over a finite set $\cs$ of $d$
elements, called the space of {\em states}.  This space may have
additional structure (e.g., it may be a graph), allowing the
definition of a notion of locality; and the set of objective functions
may have special properties depending on the particular COP. In
general multiple local
minima may be present.
Finding  a solution by exhaustive search is hard in general, due
to the large size of the search space. Therefore, more efficient
optimization approaches are desirable.  Efficiency is typically
quantified in terms of how the resources needed to find the optimum
scale with the {\em instance size}, which is typically polynomial in
$\log{d}$.

Simulated Annealing (SA) is a possible generic strategy for solving a
COP~\cite{KGV83}.  The idea of SA is to imitate the process undergone
by a metal that is heated to a high temperature and then cooled slowly
enough for thermal excitations to prevent it from getting stuck in
local minima, so that it ends up in one of its lowest-energy states.
In SA, the objective function plays the role of energy, so the lowest
energy state is the optimum.  This process can be simulated using
different techniques; we focus on discrete Markov chain Monte-Carlo
(MCMC).  These methods are often used to numerically obtain properties
of, for example, classical physical lattice systems in
equilibrium~\cite{NB99}. The general idea of MCMC is to stochastically
generate a sequence of states via a process that converges to a target
probability distribution. This is the Boltzmann distribution at the
low final temperature in the case of SA.  The efficiency of the method
relies on the fact that, in general, only a small proportion of states
contribute significantly to the determination of properties in
equilibrium.  Therefore, if a good state-generating rule is chosen,
the MCMC algorithm can explore the most relevant states only,
outperforming exhaustive search.

One way to estimate the implementation complexity of SA using MCMC is
to count the number of times that the state-generating rule must be
executed (i.e., the number of generated states) in order that the
desired distribution is reached within an acceptable error.  This
complexity, denoted by $\cn_{SA}$, is of order
$\co(\log(d/\epsilon^2)/\delta)$ (see Sec.~\ref{siman}).  Here, $\delta$
is the minimum spectral gap of the stochastic matrices used to
generate states for the COP via MCMC~\cite{St05}, while $\epsilon$ is
the error probability, that is, the probability that the final state
sampled via this process is not a solution (not in the set $\mathbb{S}_0$ of
optimal states).  
Ideally, $\cn_{SA}$ is insignificant compared to the size of
the state space.  This is the situation, for example, when computing
physical properties of the Ising spin model using the Metropolis
rule~\cite{NB99}.  In this example $\cn_{SA}$ is known to be of
order $\co(N^2)$ for a system of $N$ spins, while the state space
dimension is $d=2^N$.  Nevertheless, $\cn_{SA}$ can
increase rapidly with $N$ if the interaction strengths are made
random~\cite{Ba82}, making the problem intractable in general.
In this case, this is due to the gap $\delta$ becoming exponentially small
in $N$ (instance size).  Therefore, finding new
 methods with better scaling in $\delta$, yielding speedups
over SA, is of great importance.

Quantum mechanics provides new resources with which to attack these
optimization problems~\cite{VARIOUS1,SBO07, Farhi}.  Quantum computers
(QCs) can theoretically solve some problems, including integer number
factorization and search problems, more efficiently than today's
conventional computers~\cite{VARIOUS2}. Still, whether a QC can solve
all COPs more efficiently than its classical counterpart is an open
question. In this paper we show that QCs can also be used to speed up
the simulation of classical annealing processes. That is, we present a
new quantum algorithm that can be seen as the quantum analogue of SA
using MCMC, but for which the number of times that the
state-generating rule is called ($\cn_{QSA}$) is greatly reduced to
$\co(\log^3(d/\epsilon^2)/(\sqrt{\delta}\epsilon^2))$, to achieve
error bounded by $\epsilon$, in a single run.  This speed-up is most
significant for hard instances where $\delta \ll 1$.  Our quantum
simulated annealing algorithm (QSA) is constructed using ideas and
techniques from quantum walks~\cite{Am03,Sz04} and quantum phase
estimation~\cite{CEM98,NC00}. The QSA also exploits the so-called
quantum Zeno effect~\cite{MS77,IHB90}, in which after $Q = \cO(1/\Delta t)$
measurements of a quantum system at short time-intervals $\Delta t$
 the state is collapsed onto the ground state with total probability
$1-\co(\Delta t)$.

This paper is organized as follows.  First, in Sec.~\ref{siman}, we
describe the implementation of SA using discrete-time MCMC, and in
Appendix~\ref{appendixa} we derive a rate at which the temperature of
a classical system can be lowered to assure convergence to the set of
ground states.  To do this we adapt the results obtained for the
continuous-time case in Ref.~\cite{St05}.  The rate that we obtain is
similar to the one in Ref.~\cite{GG84} for those cases where $\delta$
decreases exponentially with the problem size (cf. Ref.~\cite{SBO07}).
In Sec.~\ref{qwalks} we describe a quantization of a reversible Markov
chain in terms of quantum walks. Our quantization is a
similarity-transformed version of the one used in
Refs.~\cite{Sz04,MNR07} to speed up search problems.  It constructs,
from the transition matrix of the Markov chain, a unitary operator
acting on a set of quantum states corresponding to the classical ones.
In Sec.~\ref{qalg} we describe our QSA and obtain the corresponding
implementation complexity, exhibiting a quantum speed-up with respect
to classical SA. Since our QSA makes calls to the phase estimation
algorithm, we describe phase estimation in Appendix~\ref{appendixb}.
Finally, we present the conclusions in Sec~\ref{concl}.


\section{Simulated Annealing and Monte-Carlo Techniques for Markov Processes}
\label{siman}
We consider the simulation of a classical annealing
process via MCMC, and give annealing rates such that the final sampled
state is almost certain to be in the set $\mathbb{S}_0$ of optimal
solutions to a COP.  To do this, we first need a formulation of the
COP in terms of an equivalent problem in which $\mathbb{S}_0$ consists
of the states that minimize some real-valued cost function $E$ on the
state space. Usually, $E$ is regarded as the energy function of a
classical system $\cs$, so the optimal solutions to the COP are
represented by the ground states of $\cs$.  For concreteness, we sometimes
think of $\cs$ as defined on a lattice with $N$ vertices, having a finite
state space $\{\sigma \}$ of size $d=\co(\exp(N))$.

A ground state can be reached by annealing slowly enough, 
starting with $\cs$ at sufficiently high temperature. 
The MCMC simulation of this process, described in terms of the {\em inverse temperature} 
$\beta \equiv 1/T$,
begins by sampling a state $\sigma^{(0)}$ from the uniform
distribution.  The annealing process is determined by a choice of an
{\em annealing schedule}, i.e. a finite increasing sequence $\beta_1 <
\beta_2< \ldots <\beta_P$, and by a sequence of {\em transition rules}
$\{ M(\beta_k)\}$.  Each $M(\beta_k)$ is a stochastic matrix whose
elements $m_{\sigma \sigma'} (\beta_k)$ are transition probabilities
from $\sigma$ to $\sigma'$.  $M(\beta_k)$ is chosen to have the
Boltzmann distribution at $\beta_k$ as its unique equilibrium
distribution.

At each step $k$,
a new state $\sigma^{(k)}$
is stochastically generated from $\sigma^{(k-1)}$ according to the
transition probabilities $M(\beta_k)$.
The annealing schedule is chosen to
give an acceptable upper bound $\epsilon$ on the probability of error
(of not ending up in $\mathbb{S}_0$).  For simplicity, we consider an
annealing schedule such that $\Delta \beta = \beta_k -\beta_{k-1} \ll
1$ is constant, and thus $\beta_f \equiv P \Delta \beta$.  In general
the annealing schedule may strongly depend on $\beta_k$.  In our
case the overall implementation complexity of the algorithm with constant
$\Delta \beta$ is of the same order as for a general annealing
schedule, so the analysis below is valid for both situations.

We choose $\Delta \beta =\co( \delta/E_M)$, where $\delta$ is the
minimum spectral gap of the matrices $M(\beta_k)$ at inverse
temperature $\beta_k = k \Delta \beta$, and $E_M := \max_\sigma
|E[\sigma]|$.  In Appendix~\ref{appendixa} we show that for $\beta_f
=\co( \gamma^{-1} \log(d/\epsilon^2))$, the probability of not ending
in a solution is no greater than $\epsilon$ [see Eq.~(\ref{saerror})].
$\gamma$ is the spectral gap of $E$.  The implementation
complexity  of SA is then given by $P=\beta_f/\Delta \beta$.  We obtain
\begin{equation}
\label{nsa}
\cn_{SA} =\co(\beta_f E_M/ \delta) = \co\(\frac{E_M}{\gamma}\frac{ \log(d/\epsilon^2)}{ \delta} \)
\end{equation} 
for a success probability greater than $1-\epsilon$. 
The dependence of $\cn_{SA}$ on $\delta^{-1}$ is
characteristic of Markov processes and, although Eq.~(\ref{nsa}) only gives 
an upper bound on the resources required for the implementation of SA,
such a dependence on the spectral gap may be unavoidable~\cite{Al81}.

Remarkably, a similar algorithm implemented on a {\em quantum}
computer has a reduced implementation complexity for those
hard instances where $\delta \ll 1$. This is described in the
following sections.


\section{Quantum Walks and Ergodic Markov Chains}
\label{qwalks}
Discrete-time quantum walks were introduced as the quantum analogues of classical
random walks~\cite{ABN01,Ke03}. 
Here, we focus on those bipartite quantum walks defined in
Refs.~\cite{Sz04,MNR07} for the purpose of obtaining
quantum speed-ups in search problems.  
Such quantum walks,  
which we describe below, can also be derived from 
Ref.~\cite{Am03}.

To define the bipartite quantum walk, we first associate each
classical state $\sigma$ of $\cs$ with a quantum state $\ket{\sigma}$
of an orthonormal basis of a $d$-dimensional Hilbert space $\ch$. We
then consider a tensor product
Hilbert space $\ch_A \otimes \ch_B$ of two copies of $\ch$.  As in SA,
we assume a given stochastic matrix $M(\beta)$ describing the Markov
process in $\cs$, with $M(\beta)$ satisfying the detailed balance
condition: $\pi^\sigma m_{\sigma \sigma'}= \pi^{\sigma '}
m_{\sigma ' \sigma}$, with $\pi^\sigma=e^{-\beta E[\sigma]}/\cz$ the
components of the equilibrium distribution ($\cz=\sum_\sigma e^{- \beta E[\sigma]}$ is the 
partition function).  In the
following we omit the dependence on $\beta$ unless necessary.  We
define isometries $X$ and $Y$ that map states of $\ch$ to states of
$\ch_A \otimes \ch_B$ as
\begin{align}
X \ket{\sigma} &= \ket{\sigma} \sum_{\sigma'} \sqrt{m_{\sigma\sigma'}} \ket{\sigma'}, \\
Y \ket{\sigma'} &= \sum_{\sigma} \sqrt{m_{\sigma'\sigma}} \ket{\sigma}\ket{\sigma'} \;.
\end{align}
The symmetric operator $H=X^\dagger Y $, acting on $\ch$, has elements
$h_{\sigma \sigma'}=\sqrt{m_{\sigma \sigma'} m_{\sigma'
    \sigma}}$~\cite{Sz04}. Because of detailed balance, we can write
$H \equiv e^{\beta H_c/2} M e^{-\beta H_c/2}$, with $H_c$ the diagonal
operator $ H_c \ket{\sigma}=E[\sigma] \ket{\sigma}$.
Therefore, the eigenvalues
$\lambda_0=1>\lambda_1\ge \cdots \ge \lambda_{d-1} \ge 0$ of $H$ are
those of $M$.  If $\ket{\phi_j}$ denotes the eigenstate of $H$ with
eigenvalue $\lambda_j$, we have for $j=0$~\cite{SBO07}
\begin{equation}
\label{qstate2}
\ket{\phi_0} \equiv \sum_{\sigma} \sqrt{\pi^\sigma} \ket{\sigma}\equiv \frac{e^{-\beta H_c/2}}{\sqrt{\cz}} \sum_\sigma \ket{\sigma}.
\end{equation}

The isometries $X$ and $Y$ define unitary operators $U_X$ and $U_Y$,
acting on $\ch_A \otimes \ch_B$, via
\begin{eqnarray}
\label{ux}
 U_X \ket{\sigma\,\mathfrak{0}} &\equiv& X\ket{ \sigma},\\
 \label{uy}
U_Y \ket{\mathfrak{0}\,\sigma} &\equiv& Y \ket{\sigma} \;,
\end{eqnarray}
with $\ket {\mathfrak{0}}$ a selected state in $\ch$.
The action of $U_X$ and $U_Y$ in the remaining subspace is irrelevant.
We now define $R_1$ to be the reflection operator through the subspace
spanned by $\{\ket{\sigma\, \mathfrak{0}}\}$ and $R_2$ the reflection
operator through the subspace spanned by $\{U_X^\dagger U^{\;}_Y \ket{
  \mathfrak{0}\,\sigma}\}$. Thus,
\begin{align}
R_1&\equiv 2 \Pi_1 - \one\otimes\one, \\
R_2&\equiv 2  \Pi_2 - \one\otimes\one\;,
\end{align}
where $\Pi_1$ and $\Pi_2$ are the projectors
\begin{align}
  \Pi_1 &\equiv \one \otimes \ket{\mathfrak{0}} \bra{\mathfrak{0}} \ , \\
  \Pi_2 &\equiv U_X^\dagger U^{\;}_Y (\ket{\mathfrak{0}} \bra{\mathfrak{0}} \otimes \one )U_Y^\dagger U_X^{\;} \ .
  \end{align}
The unitary operation (rotation) $W(M) \equiv R_2 R_1$ defines the
bipartite quantum walk based on the Markov chain
$M$.  This walk is related to the one used in Refs.~\cite{Sz04,MNR07} 
by a unitary, but $\beta$-dependent, 
similarity transformation;  using the transformed version is necessary
for our QSA to work.

The spectrum of $W(M)$ can be directly related to the spectrum of
$M$~\cite{Sz04}. Defining the phases $\varphi_j\equiv \arccos
\lambda_j$, so that
\begin{align}
 H \ket{\phi_j} &= \cos \varphi_j  \ket{\phi_j}  = X^\dagger Y \ket{\phi_j}\;,
\end{align}
we have $\varphi_0=0$. When $\varphi_1 \ll 1$, the
spectral gap of $M$ (or $H$) is $1-\lambda_1 \approx
(\varphi_1)^2/2$. From Eqs.~(\ref{ux}) and~(\ref{uy}),
\begin{align}
  \Pi_1 U_X^\dagger U_Y^{\;} \ket{\mathfrak 0\,\phi_j} &= \cos{\varphi_j} \ket{\phi_j\, \mathfrak 0} \\
  \Pi_2 \ket{\phi_j\,\mathfrak 0} &= \cos{\varphi_j}\, U_X^\dagger U_Y^{\;} \ket{\mathfrak 0\,\phi_j} \;,
\end{align}
so the action of $W(M)$ in the (at most) two-dimensional subspace
spanned by $\{\ket{\phi_j\, \mathfrak{0}}, U_X^\dagger U_Y^{\;}
\ket{\mathfrak{0}\,\phi_j}\}$ is an overall $4\varphi_j$ rotation
along an axis perpendicular to that subspace~\cite{KOS07}.  Thus the
eigenphases of $W(M)$ are $\pm 2 \varphi_j$, and its eigenvectors for
$j \ne 0$ are:
\begin{align}
\label{wvector}
\ket{\psi_{\pm j}} = \frac {\pm i } {\sqrt 2 \sin{\varphi_j}} 
\( e^{\mp i \varphi_j}\ket{\phi_j\, \mathfrak{0}} -  U_X^\dagger U_Y^{\;} 
\ket {\mathfrak{0}\, \phi_j}\)\;.
\end{align}
When $j=0$, we have
\begin{equation}
\label{qstate3}
\ket{\psi_0} \equiv \ket{\phi_0 \ \mathfrak{0}},
\end{equation}
so a quantum algorithm that prepares the {\it quantum Gibbs} state
$\ket{\psi_0}$ allows us to sample from the desired (equilibrium)
distribution by measuring $\ch_A$ in the basis $\{\ket{\sigma}\}$.
All the other eigenphases of $W(M)$ that were not described are either
$0$ or $\pi$.

The (quantum) implementation complexity of $U_X$ and $U_Y$ is
proportional to the (classical) implementation complexity of a
single step of the MCMC method described in Sec.~\ref{siman}, because
$U_X^{\;}$, $U_X^\dagger$, $U_Y^{\;}$, and $U_Y^\dagger$ may be implemented 
using a {\em reversible} version of the classical algorithm that
computes a matrix element of $M$. It follows that the implementation
complexity of $W(M)$ is proportional to the classical complexity of
implementing four steps in the MCMC method.

The operations $W(M)$ will be used below to implement the QSA.  An
important property that follows from our definition of $W(M)$ is that the
overlap between the quantum Gibbs state $\ket{\psi_0(\beta)}$ and any
other eigenstate in the $0$-eigenphase subspace,
at any $\beta'$,
is zero.  To show this note that $\ket{\phi_j}$ is a complete basis
for $\ch$, and $\ket{\phi_j \, \mathfrak{0}}=\frac{1}{\sqrt{2}}
[\ket{\psi_{+j}} + \ket{\psi_{-j}}]$ ($j \neq 0$).  Thus,
\begin{eqnarray}
\label{eq:0t}
\ket{\psi_0(\beta)}&=&\sum_{j=0}^{d-1} c_j \ket{\phi_j(\beta')\,\mathfrak{0}} \\
\nonumber
&=&c_0 \ket{\psi_0(\beta')}+ \sum_{j=1}^{d-1} \frac{c_j}{\sqrt{2}} 
[ \ket{\psi_{+j}(\beta')}  +\ket{\psi_{-j}(\beta')}]\;.
\end{eqnarray}
Our algorithm uses this property to keep the state
$\ket{\psi_0(\beta)}$ separated form the remaining degenerate
subspace.


\section{Quantum Simulated Annealing Algorithm}
\label{qalg}
The QSA that we propose is basically a sequence of phase estimation
algorithms (PEAs) projecting onto the quantum Gibbs state that is
associated with the equilibrium state of $\cs$ for different
temperatures. The implementation complexity of SA is dominated by the
gap of the stochastic matrix, which constrains the annealing
schedule. For the QSA algorithm, the total implementation complexity
is dominated by the implementation complexity of each PEA, given by
the eigenphase gap of the quantum walk. Because the latter is
(quadratically) larger than the former, the QSA algorithm results in a
(quadratic) quantum speed-up of SA.

We consider a sequence of inverse temperatures $\{\beta_k = k \Delta
\beta\}$, with $k=1,\ldots,Q$, and $\beta_f=\beta_Q=Q\Delta
\beta$. The choice of $\Delta \beta$ differs from the one used for
SA. To understand the QSA, we begin by performing a Taylor series
expansion of $\ket{ \phi_0(\beta_{k-1})}$ [Eq.~(\ref{qstate2})] in
$\beta_{k}$. We obtain,
\begin{align}
\label{expans1}
\ket{ \phi_0(\beta_{k-1})}& = \left( 1 - \frac{ \Delta \beta }{2}\(
\langle E\rangle _{\beta_{k}}-H_c\) \right) \ket{\phi_0(\beta_{k})
} \nonumber \\ &\qquad+\cO( \nu^2 )  \;,
\end{align}
where $\langle E \rangle _{\beta_{k}}=\sum_\sigma E[\sigma]
e^{-(\beta_{k} ) E[\sigma]}/\cz (\beta_k) \equiv
\bra{\phi_0(\beta_{k})} H_c \ket{\phi_0(\beta_{k})}$ is the
expectation value of the energy (cost function), and $\nu = \Delta
\beta\, E_M$.  The (squared) overlap for two adjacent values of $\beta$ is
\begin{equation}
\label{zeno}
|\bra{ \phi_0 (\beta_{k})} \phi_0(\beta_{k-1})\rangle|^2 = 1  -\cO({\nu}^2) \;.
\end{equation}
It follows that the probability of successful preparation of
$\ket{\phi_0(\beta_f)}$, after $Q=\co(1/\nu)$ projective measurements,
can be bounded below by $1-\co(\nu)$. This is called the quantum Zeno
effect~\cite{MS77,IHB90}.  Our QSA algorithm performs such projections
by calling the PEA at $\beta_1,\ldots,\beta_f$. This technique was
used in Ref.~\cite{CDF02} to obtain the quadratic quantum speed-up for
Grover's unstructured search problem.

\begin{figure}
\includegraphics[width=3.3in]{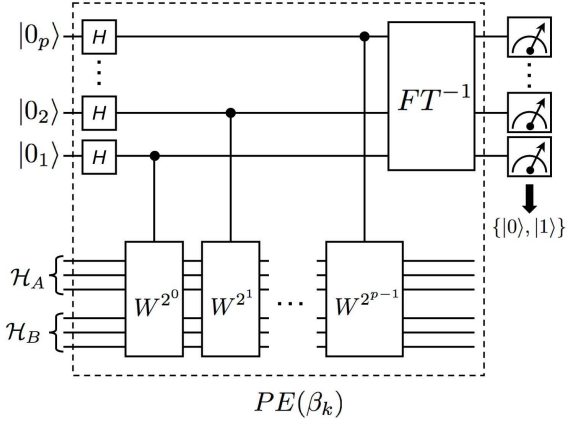}
\caption{Phase estimation algorithm (subroutine) for the quantum
simulated annealing algorithm. The first register of $p$ qubits is
used to encode the eigenphases of $ W(M(\beta_{k}))$. The second
register denotes the bipartite system $\ch_A \otimes \ch_B$.  The
algorithm takes as input, in the second register, a quantum state sufficiently
close to $\ket{\psi_0(\beta_{k-1})}$.  A sequence of controlled
$W(M(\beta_{k+1}))$ operations is performed and the inverse of the
quantum Fourier transform is then applied; the composition of all
these unitary operations is denoted $PE(\beta_{k})$.  Finally, the
first register is measured.  When the result of the measurement is
such that the first register is projected onto $\ket{0}=\ket{0_1\ldots
0_p}$, the PEA outputs a state close to $\ket{\psi_0(\beta_{k})}$ in
the second register. }
\label{peafig}
\end{figure}

The PEA at the $k$th step is depicted in Fig.~\ref{peafig}. The
$p$ ancillary qubits composing the first register are used to encode
the eigenphases of $W(M(\beta_{k}))$ as binary fractions. In
particular, $2 \varphi_0 =0 = [0_1 \ldots 0_p]_2$. The integer
$p$ is chosen to satisfy $2^p =\co(1/(\nu \sqrt{\delta}))$. This
choice allows us to bound the error due to the impossibility of
representing the phases $2\varphi_j$ with $p$ bits (see
Appendix~\ref{appendixb} and Ref.~\cite{CEM98}).  The PEA gets as
input a state close to $\ket{0 \, \psi_0(\beta_{k-1})}$.  It starts with a
sequence of unitary gates that includes $2^p-1$ actions of the
operation $cW(M) =cR_2 cR_1= U_X^\dagger U_Y^{\;} cP_{\mathfrak{0}_A}
U_Y^\dagger U_X^{\;} cP_{\mathfrak{0}_B}$, controlled on the states
$\ket{1_i}$ of the first register ($i=1,\ldots,p$). Here,
$cP_{\mathfrak{0}_A}$ and $cP_{\mathfrak{0}_B}$ are the controlled
selective sign change operations on the states $\ket{\mathfrak{0}}$ of
$\ch_A$ and $\ch_B$, respectively. It continues with an inverse
quantum Fourier transform, and finally the first register is measured
in the computational basis. For the given input state, the PEA outputs
a state close to $\ket{0 \, \psi_0(\beta_{k})}$ with probability
close to one.  Since each use of $cW(M(\beta_{k}))$ has complexity
proportional to that of four steps of the classical MCMC method, the
overall implementation complexity of the PEA is $\cn_{PEA}=\co(1/(\nu
\sqrt{\delta}))$.

\begin{figure}
\includegraphics[width=3.3in]{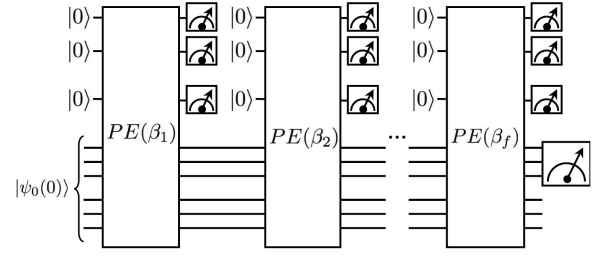}
\caption{Quantum simulated annealing algorithm. The algorithm is a
sequence of $Q$ calls to the PEA at $\beta_1,\ldots,\beta_f$. After
the last call, the state of $\ch_A \otimes \ch_B$ is
 close to $\ket{\psi_0(\beta_f)}
\equiv \ket{\phi_0(\beta_f) \, \mathfrak{0}}$, with probability close
to one. A measurement on $\ch_A$  returns a state $\sigma$ in the ground state space of $\cs$ with probability greater than $1-\epsilon$.}
\label{qsafig}
\end{figure}

The QSA is depicted in Fig.~\ref{qsafig}. It is composed of
$Q$ calls to the PEA, with a final measurement of $\ch_A$ in the
$\ket{\sigma}-$basis.  In Appendix~\ref{appendixb} we show that, after
the measurement, the probability of finding $\ch_A$ in the
excited space can be bounded as
\begin{equation}
\label{probbound2}
\cp(\sigma \not \in \mathbb{S}_0) \le d e^{-\beta_f \gamma} + \tau' Q \nu^2\;,
\end{equation}
for some constant $\tau'=\co(1)$. We seek to make the above error of order
$\epsilon$.  Choosing $\beta_f=\gamma^{-1} \log(2d/\epsilon^2)$, as in
SA, makes the first term on the right hand side of
Eq.~(\ref{probbound2}) of order $\co(\epsilon^2)$. Thus we need $\tau' Q
\nu^2= \co(\epsilon)$.  The condition $Q \Delta \beta=\beta_f$ implies
$\Delta \beta=\co(\epsilon/(\beta_f E_M^2))$ and $Q=\co((\beta_f
E_M)^2/\epsilon)$. Finally, because $\cn_{QSA}=\co(Q \cn_{PEA})$, we
obtain
\begin{equation}
\cn_{QSA}=\co\(\frac{(\beta_f E_M)^3}{\epsilon^2 \sqrt{\delta}}\)
=\co\( \(\frac{E_M}{\gamma}\)^3 \frac{\log^3(2d /\epsilon^2)}{\epsilon^2 \sqrt{\delta}}\).
\end{equation}

The above scaling with $1/\epsilon^2$ is for a single run of the QSA.  
Typically,
repetition of the QSA makes the error exponentially low
in the amount of resources used, so the dependence of $\cn_{QSA}$ 
on $\epsilon$ can be made logarithmic.
The cubic scaling with 
the parameter $E_M/\gamma$ is also worse than classical SA's linear scaling, 
but this is relatively unimportant as in most applications this 
parameter will be bounded by a constant or a polynomial in instance size. 

Note that, since only the state of $\ch_A$ is important for our purposes , the QSA can be implemented without measuring the ancillary qubits used in each PEA. In this case, the operations $FT^{-1}$ can be avoided~\cite{CDF02}. This is because the quantum Zeno effect relies on the {\em decoherence} introduced by the interaction with the ancillae, not the measurement itself.

\section{Conclusions}
\label{concl}
We have presented a quantum algorithm to simulate classical annealing
processes by quantization of the simulated annealing algorithm
implemented with MCMC methods. Such a quantization has been done by
using techniques borrowed from quantum walks and quantum phase
estimation. Our algorithm also exploits the quantum Zeno effect.  We
have shown that, if $ \epsilon$ denotes an upper bound to the
probability of not finding an optimal solution to a COP, the QSA
requires resources $\cn_{QSA} = \co\( \frac{\log^3(2d
  /\epsilon^2)}{\epsilon^2 \sqrt{\delta}}\)$, with $\delta$ the
spectral gap.  Thus QSA outperforms SA in those problems where $\delta
\ll 1$, such as finding a ground state of a spin glass.  SA requires
$\cn_{SA}=\co( \log( d/\epsilon^2)/\delta)$ to assure the same error
probability.  Even if SA could be implemented more efficiently, the
scaling of $\cn_{SA}$ with $\delta^{-1}$ may be
unavoidable~\cite{Al81}.  Since initializing with a state close to
$\ket{\phi_0(\beta_f )}$ is not required by the QSA, our result has
implications in the mixing time problem studied in Ref.~\cite{Ri07}.

We expect that similar quantum speed-ups hold for the simulation of
more general classical annealing processes. Moreover, our algorithm
can easily be extended to simulate continuous-time annealing.  Also,
by choosing $\beta_f=1/T$, with $T>0$, the QSA can be used to speed up
the calculation of finite-temperature thermodynamic properties of
classical systems on a lattice.

Finally, our QSA is one possible quantum algorithm to simulate an
annealing process. One may wonder if other quantum algorithms, based
on quantum adiabatic evolutions, can still provide similar quantum
speed-ups.  The adiabatic theorem of quantum mechanics yields similar
convergence rates. A simple, but not rigorous, proof is given by
considering the adiabatic condition (cf. \cite{Messiah}):
\begin{equation}
\label{adiab}
\partial_t \beta(t) \left|\frac{ \langle \psi_{\pm j}(\beta) | \partial_\beta \psi_0 (\beta) 
\rangle}{2\varphi_j}\right| \le \partial_t \beta(t) \frac{E_M}{2 \varphi_1} \le \epsilon,
\end{equation}
with $j \neq0$. Other $0$-eigenphase states have not been considered as they do 
not overlap with $\ket{\psi_0(\beta)}$ at first order [Eq.~(\ref{eq:0t})].
The overall implementation
complexity of the adiabatic evolution (i.e., total evolution time) determined by Eq.~(\ref{adiab})
 is $\co(1/(\epsilon \sqrt{\delta}))$.  Details will be given elsewhere.

 \acknowledgments We thank Stephen Jordan for discussions and for
 pointing out Ref.~\cite{CDF02}.  This research was supported by
 Perimeter Institute for Theoretical Physics.  Research at Perimeter
 Institute is supported by the Government of Canada through Industry
 Canada and by the Province of Ontario through the Ministry of
 Research and Innovation. This work was also carried out partially
 under the auspices of the NNSA of the US DOE at LANL under Contract
 No. DE-AC52-06NA25396 and by NSF Grant No. PHY-0653596.

\begin{appendix}
\section{Convergence of classical simulated annealing}
\label{appendixa}
We now obtain an annealing schedule that assures convergence to the
desired state when SA is implemented using discrete MCMC methods. The
following analysis is based on Ref.~\cite{St05}, where similar rates
have been obtained in the continuous-time case.  Assume that we start
with a state sampled from some probability vector $\vec
\mu(0)=\frac{1}{d}(1,\cdots,1)$ (i.e., the uniform
distribution). After $P$ steps, this state evolves to
\begin{equation}
\label{kolm}
\vec \mu(\beta_f) = (\mu^1(\beta_f),\ldots,\mu^d(\beta_f)) = \(\prod_{k=1}^P M(\beta_k)\) \vec \mu(0),
\end{equation}
with $\beta_k =k \Delta
\beta$. 
Because $M$ is stochastic, normalization is preserved: $\sum_{\sigma=1}^d \mu^\sigma(\beta_f)=1$. 
Let $\vec \pi(\beta_f)=(\pi^1(\beta_f),\ldots,\pi^d(\beta_f))$ be 
the desired (Boltzmann) equilibrium distribution after the annealing
process. That is, $M(\beta_f) \vec \pi(\beta_f) \equiv
\vec\pi(\beta_f)$, and also $\sum_{\sigma=1}^d \pi^\sigma(\beta_f)=1$.  From the
Cauchy-Schwarz inequality we obtain, for the probability of error,
\begin{eqnarray}
\label{cs1}
  \cp(\sigma^{(P)} \not\in \mathbb{S}_0) &=& \sum_{\sigma \notin \mathbb{S}_0} \mu^\sigma(\beta_f) \\
   \nonumber
  &\le& \sqrt{ \left[ \sum_{\sigma=1}^d \frac {\(\mu^\sigma(\beta_f)\)^2} {\pi^\sigma(\beta_f)} \right] \left[  \sum_{\sigma \notin \mathbb{S}_0} \pi^\sigma(\beta_f) \right]} .
\end{eqnarray}
Considering the worst case, in which all non-ground states have energy 
$E[\mathbb{S}_0]+\gamma$ gives: 
\begin{align}
\label{boundsa1}
  \sqrt{ \sum_{\sigma \notin \mathbb{S}_0} \pi^\sigma(\beta_f)} \le \sqrt{d}\, e^{- \beta_f \gamma/2} \;,
\end{align}
where $\gamma=\min_{\sigma \notin \mathbb{S}_0}
|E[\sigma]-E[\mathbb{S}_0]|$ is the spectral gap of $E$ and $d$ is the
dimension of the state space $\cs$. Equation~(\ref{boundsa1}) was
obtained considering the worst case scenario in which the space of
states having energy $E[\mathbb{S}_0] + \gamma$ is highly
degenerate. Thus
\begin{align}
\label{eq:sap}
  \cp(\sigma^{(P)} \not\in \mathbb{S}_0)  \le \sqrt{d}\, e^{- \beta_f \gamma/2} \| \vec h(\beta_f)\|_2\;,
\end{align}
where $\| \vec h(\beta_f)\|_2$ denotes the $2$-norm of
\begin{align}
  \vec h(\beta_f) \equiv \(\frac{\mu^1(\beta_f)}{\sqrt {\pi^1(\beta_f)}},\ldots,\frac{\mu^d(\beta_f)}{\sqrt {\pi^d(\beta_f)}}\)\;.
\end{align}

To bound $\| \vec h(\beta_f)\|_2$, we define, as in Sec. III, the
symmetric matrix $H(\beta_k) \equiv e^{\beta_k H_c/2} M(\beta_k)
e^{-\beta_k H_c/2}$, with $H_c$ the diagonal matrix having $E[1],\ldots,E[d]$
as elements.  We denote by
$\lambda_1(\beta_k)=1>\lambda_2(\beta_k)\ge \cdots \ge
\lambda_d(\beta_k)\ge 0$ the eigenvalues of $M(\beta_k)$ and
$H(\beta_k)$.  The eigenvector of $H(\beta_k)$ with largest eigenvalue
is~\cite{SBO07}
\begin{align}
\label{qstate}
\sqrt {\vec \pi(\beta_k)} &= \(\sqrt{\pi^1(\beta_k)},\ldots,\sqrt{\pi^d (\beta_k)}\) \nonumber \\
&\equiv \frac{1}{\sqrt{\cz}} \(e^{-\beta_k E[1]/2},\ldots,e^{-\beta_k E[d]/2}\)\;,
\end{align}
where $\cz=\sum_{\sigma=1}^d e^{-\beta_k E[\sigma]}$ is the partition
function. Denote now as $\delta=\min_k\{1-\lambda_2(\beta_k)\}$
the minimum spectral gap of the matrices $H(\beta_k)$ (or
$M(\beta_k)$).  We will show that, when $\delta \ll 1$,
an annealing rate $\Delta \beta$ satisfying 
\begin{equation}
\label{rate1}
\Delta \beta E_M \le \tau \delta,
\end{equation}
implies $\|\vec h(\beta_f)\|_2 \le \sqrt 2$~\cite{comment1}. Here, $E_M =\max_\sigma
|E[\sigma]| $ and $\tau$ is a $\co(1)$ constant.

We start by writing
\begin{align}
\label{evst}
  \Delta \vec \mu(\beta_k) &\equiv \vec \mu(\beta_{k+1}) - \vec \mu(\beta_k)  \\
  \nonumber
  &= \(M(\beta_{k+1}) - \one \) \vec \mu(\beta_k)\;,
\end{align}
where $\vec \mu(\beta_k) = \prod_{k'=1}^k M(\beta_{k'}) \vec \mu(0)$.
Also, from the Taylor series expansion of $\vec \pi(\beta_k)$ and using Eq.~(\ref{rate1}), we obtain
\begin{align}
\label{eqdist}
  &\sqrt{\(\vec \pi(\beta_{k+1})\) }- \sqrt{ \(\vec \pi(\beta_k)\)}=\\  \nonumber 
  &\quad =  \frac 1 2 \Delta \beta (\langle E \rangle_{\beta_k} -H_c) \sqrt{\(\vec \pi(\beta_k)\)} +\cO(\delta^2)\;,
\end{align}
where $\langle E \rangle_{\beta_k}=\sum_{\sigma=1}^d E[\sigma] e^{-\beta_k E[\sigma]}/\cz$ is the
expectation value of $E$ at $\beta_k$.
Combining Eqs.~(\ref{evst}) and~(\ref{eqdist}), and defining $\vec h(\beta_k) =\left(\frac{\mu^1(\beta_k)}{\sqrt{\pi^1(\beta_k)}},\ldots,\frac{\mu^d(\beta_k)}{\sqrt{\pi^d(\beta_k)}} \right)$, we have
\begin{align}\label{eq:dh}
  \Delta& \vec h(\beta_k) \equiv  \vec h(\beta_{k+1}) - \vec h(\beta_k) \nonumber \\
  &=  (H(\beta_{k+1} ) - \one)\vec h(\beta_k)\(1+\cO(\delta)\) \nonumber \\
  &\quad -\(\frac 1 2 \Delta \beta (\langle E \rangle_{\beta_k} -H_c) \)  \vec h(\beta_k) + \cO(\delta^2)\;.
\end{align}
Therefore, if  $\langle \cdot\ , \cdot \rangle $ refers to the standard inner product, 
\begin{align}
  \langle \vec h(\beta_k),\,& \Delta \vec h(\beta_k) \rangle = \nonumber \\
  &=  \langle \vec h(\beta_k), (H(\beta_{k+1})-\one)\vec h(\beta_k) \rangle (1+\cO(\delta)) \nonumber \\
  \label{sabound4}
  &  -\frac 1 2 \Delta \beta \langle \vec h(\beta_k), (\langle E \rangle_{\beta_k} -H_c) \vec h(\beta_k)\rangle+ \cO(\delta^2)\;.
\end{align}

The first term in Eq.~(\ref{sabound4}) can be bounded by expanding $\vec h(\beta_k)$ as a sum of
the eigenvectors of $H(\beta_{k+1})$,  denoted as $\{\vec
e_j(\beta_{k+1})\}$, with $\vec e_1(\beta_{k+1}) \equiv \sqrt { \vec
  \pi(\beta_{k+1})}$ [see Eq.~(\ref{qstate})].  Then,
\begin{align}
     \langle \vec h(\beta_k), (H(\beta_{k+1})&-\one)\vec h(\beta_k) \rangle (1+\cO(\delta))  \nonumber \\ & \le  - \delta \(\| \vec h(\beta_k) \|_2^2  -1\) +\cO(\delta^2)\;.
\end{align} 
This results in
\begin{align}
   \langle &\vec h(\beta_k),\, \Delta \vec h(\beta_k) \rangle  \nonumber \\
   &\le \(-\delta + \frac 1 2 \Delta \beta  E_M\) \|\vec h(\beta_k)\|_2^2 + \delta + \cO(\delta^2)\;,
\end{align}
where we considered that $\langle \vec h(\beta_k),H_c \vec h(\beta_k)\rangle \le E_M \| \vec h(\beta_k) \|_2^2$ and, with no loss of generality, $\langle E \rangle_{\beta_k} \ge 0$.
Therefore, the increment on $\| \vec h(\beta_k)\|_2$ can bounded as
\begin{align}
  \Delta \| \vec h(\beta_k)\|_2^2 & \equiv \|\vec h(\beta_{k+1})\|_2^2 - \|\vec h(\beta_k)\|_2^2 \nonumber \\
  &= 2 \langle \vec h(\beta_k),\Delta \vec h(\beta_k) \rangle + \| \Delta \vec h(\beta_k) \|_2^2 \\
  &\le \(-2 \delta +  \Delta \beta  E_M\) \|\vec h(\beta_k)\|_2^2 \nonumber \\
   &\quad \quad \quad \quad+ 2 \delta + \cO(\delta^2)\;.
\end{align}
Since $\|\vec h(\beta_k)\|_2^2 \ge 1$ we have, for a proper choice of $\tau=\co(1)$ in Eq.~(\ref{rate1}),
\begin{align}
\label{deltanorm}
  \Delta \|\vec h(\beta_k)\|_2^2 \le -\delta \|\vec h(\beta_k)\|_2^2 + 2 \delta\;.
\end{align}
Equivalently,
\begin{align}
\label{deltanorm2}
 \|\vec h(\beta_{k+1})\|_2^2   \le (1-\delta) \|\vec h(\beta_k)\|_2^2 + 2 \delta\;.
\end{align}
Furthermore, the condition $\vec\pi(0) \equiv \vec \mu(0)$ yields to
$\|\vec h(0)\|_2 = 1$. Iterating Eq.~(\ref{deltanorm2}) for $k'=0,\ldots,k$, we obtain
\begin{align}
  \|\vec h(\beta_k)\|_2^2 \le 2 - (1-\delta)^{k} \le 2\;.
\end{align}
Finally, using Eq.~\eqref{eq:sap}, we obtain the desired bound on the probability of error, given by
\begin{align}
\label{saerror}
  \cp(\sigma^{(P)} \not\in \mathbb{S}_0)  \le \sqrt{2\, d}\, e^{- \beta_f \gamma/2} \;.
\end{align}

\section{Implementation complexity of the quantum simulated annealing algorithm}
\label{appendixb}
We first show how the PEA works for the eigenphases $\pm 2\varphi_j$
of $ W(M)$, with $\varphi_0=0 < \varphi_1 \le \cdots \le \varphi_{d-1}
\le \pi/2$. We write
\begin{equation}
\label{bindec}
2 \varphi_j = 2 \pi( [.a_1^j \ldots a_p^j]_2 + \zeta_j )\equiv 2 \pi(\sum_{i=1}^p a_i^j /2^i+\zeta_j),
\end{equation}
with $|\zeta_j | \le 1/2^{p+1}$ and $2 \pi [.a_1^j \ldots a_p^j]_2$
the best $p$-bit approximation to $2 \varphi_j$.
The PEA (Fig.~\ref{peafig}) begins by applying a set of Hadamard gates
to the $p$ qubits in the first register, initialized in the state
$\ket{0}=\ket{0_1\ldots 0_p}$. These qubits are used to encode the
eigenphases as binary fractions at the end of the PEA.  The PEA then
applies a set of operations ${W^{2^{i-1}}(M)}$, with $i=1,\ldots,p$,
controlled on the states $\ket{1_i}$ of the first register.
Consider the case where the initial state of $\ch_A \otimes
\ch_B$ is one of the eigenstates $\ket{\psi_{\pm j}}$ of $ W(M)$
[Eqs.~(\ref{wvector}) and~(\ref{qstate3})]. The evolved joint state is
\begin{align}
\nonumber
\frac{1}{\sqrt{2^p}} &(\ket{0_1} + e^{\pm i  2^{0} (2 \varphi_j)} \ket{1_1}) \cdots \\
&\cdots (\ket{0_p} + e^{\pm i  2^{p-1} (2 \varphi_j)} \ket{1_p}) \ket{ \psi_{\pm j}}\;.
\end{align}
The next step is to apply the inverse of the quantum Fourier
transform, denoted by $FT^{-1}$ in Fig.~\ref{peafig}, to the first
register. Its action is given by
\begin{equation}
FT^{-1}\ket{m}  = \frac{1}{\sqrt{2^p}} \sum_{m'=0}^{2^p-1} e^{-i2 \pi m m' / 2^p} \ket{m'},
\end{equation}
where $m,m' \in [0,\ldots,2^p-1]$ are natural numbers whose binary representation denotes the states of qubits $1,\ldots,p$. The evolved (joint) state is now
\begin{equation}
  \ket{\eta}=\frac{1}{2^p} \sum_{m=0,m'=0}^{2^p-1} e^{-i2 \pi m m' / 2^p} e^{\pm i  m' (2 \varphi_j)} \ket{m\, \psi_{\pm j}}.
\end{equation}
The final step of the PEA is to perform a projective measurement of
the first register in the (computational)
$\{\ket{0_i},\ket{1_i}\}-$basis ($i=1,\ldots,p$). The probability of
projecting the first register onto some state $\ket{m}$ is determined
by $|o_{\pm j,m} |^2$, with
\begin{eqnarray}
\nonumber
o_{\pm j,m}&\equiv& \bra{m\, \psi_j}\eta \rangle \\
\nonumber
& = &\frac{1}{2^p}\sum_{m'=0}^{2^p-1} e^{-i2 \pi m m' / 2^p} e^{i  m' (2 \varphi_j)} \\
\label{ovlp}
&=& \frac{1}{2^p}\frac{1-e^{i  [2^p(2\varphi_j )-2\pi m]}} {1-e^{i (2\varphi_j-2 \pi m/2^p)}}.
\end{eqnarray}
In particular, $o_{0,m}=\delta_{0,m}$ and, since $|1-e^{ix}| \ge
2|x|/\pi$, we have $|o_{\pm j,m=0}|\le \pi /(2^p(2 \varphi_j)) $. The
error is due to the fact that, in general, $2\varphi_j$ does not admit
an exact representation using $p$ bits.

Clearly, the implementation complexity $\cn_{PEA}$ of the PEA is of
order $\co(2^p)$. The choice of $p$ depends on the overall probability
of error of the QSA. Below we show that, by choosing $|o_{\pm j,m=0}|
=\cO( \nu )$, with $\nu=\Delta \beta E_M$, the QSA is guaranteed to succeed with a probability of
error of order $\co(\epsilon)$. Furthermore, since $\min_{j,\beta}\{ \varphi_j(\beta) \}
=\co(\sqrt{\delta})$, where $\delta$ is the minimum spectral gap of
$M(\beta)$, it is enough to choose $p$ such that $2^p = \cO(1/(\nu
\sqrt \delta))$, giving a implementation complexity for each phase
estimation $\cn_{PEA}= \co( 1/(\nu \sqrt{\delta}))$.

To obtain the implementation complexity of the QSA, it is helpful to
consider the equivalent case where non of the measurements are
actually performed until after the final PEA~\cite{CDF02}.  The input
state to the first PEA is $\ket{0_\mathfrak{1} \, \psi_0(0) }$, where
we introduce the subscripts $\mathfrak{1},\ldots,\mathfrak{q}$ to
denote the sets of $p$ qubits used as ancillae in each  PEA. The
first PEA is performed at inverse temperature $\beta_1$. From
Eq.~(\ref{expans1})
\begin{equation}
\ket{0_\mathfrak{1} \, \psi_0(0) } =\( {1-\cO(\nu^2)}\)\ket{0_\mathfrak{1} \, \psi_0(\beta_1) }+
\cO(\nu) \ket{0_\mathfrak{1} \, \psi^\perp_0(\beta_1) }.
\end{equation}
Also [Eq.~(\ref{eq:0t})],
\begin{equation}
\ket{\psi^\perp_0(\beta_1)}=\sum_{j=1}^{d-1} \frac{e_j}{\sqrt{2}} [\ket{\psi_{+j}(\beta_1)}+\ket{\psi_{-j}(\beta_1)}].
\end{equation}
After the implementation of the unitary $PE(\beta_1)$ (see Fig.~\ref{qsafig}), the evolved state is
\begin{align}
\label{evol1}
&\(1-\cO(\nu^2)\)\ket{0_\mathfrak{1} \, \psi_0(\beta_1) } \\ 
\nonumber
&+\cO(\nu) \sum_{j,m} \frac{e_j }{\sqrt{2}} [o_{+j,m}\ket{m_\mathfrak{1} \, \psi_{+j}(\beta_1)}+o_{-j,m}\ket{m_{\mathfrak{1}} \,\psi_{-j}(\beta_1)}].
\end{align}
Since only the states with $m_\mathfrak{1}=0$ in the above sum contribute to the final probability of projecting onto $\ket{0_\mathfrak{1}}$ at the end of the first PEA, it is convenient to rewrite Eq.~(\ref{evol1}) as
\begin{align}
\label{evol2}
\(1-\cO(\nu^2)\)\ket{0_\mathfrak{1} \, \psi_0(\beta_1) }& + \cO(\nu^2)  \ket{0_\mathfrak{1} \, \psi_0^\perp(\beta_1)} \nonumber \\ &+ \cO(\nu) \ket{\chi_1}\;.
\end{align}
Here, $\langle \psi_0(\beta_1) \ket {  \psi^\perp_0(\beta_1)} =
\langle 0_\mathfrak{1} \ket{\chi_1} =0$ and the order of the second
term follows from the previous choice of $p$ so that $|o_{\pm j,m=0}| = \cO(
\nu)$.

We now introduce the state $\ket{0_\mathfrak{2}}$ for the second set
of $p$ qubits, and evolve with the action of $PE(\beta_2)$. 
The output of the second phase estimation gives [Eq.~(\ref{expans1})]
\begin{align}
  &\(1-\cO(\nu^2)\)\ket{0_\mathfrak{2}0_\mathfrak{1} \, \psi_0(\beta_1) } + \cO(\nu^2)  \ket{0_\mathfrak{2} 0_\mathfrak{1} \,  \psi_0^\perp(\beta_2)} \nonumber \\
  &\quad+ \cO(\nu^2)  PE(\beta_2)   \ket{0_\mathfrak{2} 0_\mathfrak{1} \,  \psi_0^\perp(\beta_1)} + \cO(\nu) \ket {\chi_2}\;,
\end{align}
with $\bra{0_{\mathfrak 1}0_{\mathfrak 2}} \chi_2 \rangle = 0$.

We repeat this procedure by introducing the states
$\ket{0_\mathfrak{3}},\ldots,\ket{0_\mathfrak{q}}$ and by evolving
with $PE(\beta_3),\ldots,PE(\beta_Q=\beta_f)$, respectively.  Denote
by $\ket{\xi}$ the evolved (joint) state of all the registers
$\mathfrak{1},\ldots,\mathfrak{q}$ and $\ch_A \otimes \ch_B$.  After
the measurement on $\mathfrak{1},\ldots , \mathfrak{q}$, the
probability of projecting onto $\ket{0_\mathfrak{q} \ldots
  0\mathfrak{1}}$ is given by $\cp_0=\bra {\xi} P_0 \ket{\xi}$, with
$P_0= \ket{0_\mathfrak{q} \ldots 0_\mathfrak{1}} \bra{0_\mathfrak{q}
  \ldots 0_\mathfrak{1}}$ the projector onto the corresponding
subspace. By a similar analysis as the ones performed above for the
first two steps, we obtain
\begin{align}
\label{0state}
&P_0 \ket{\xi} \equiv \( 1-\cO(\nu^2) \)^{Q} \ket{0_\mathfrak{q}\ldots 0_\mathfrak{1} \, \psi_0(\beta_f)} + \\
\nonumber & \cO({\nu}^2)  P_0 \sum_{i=0}^{Q-1} PE(\beta_Q)\cdots
  PE(\beta_{Q-i+1}) \ket{0_\mathfrak{q}\ldots 0_\mathfrak{1} \, 
    \psi^\perp_0(\beta_{Q-i})}.
\end{align}
Thus the probability of $\ch_A \otimes \ch_B$ being in the desired
state $\ket{\psi_0(\beta_f)}$ can be bounded below, by using
Eq.~(\ref{0state}), as
\begin{align}
\label{prob0}
\cp_0 &\ge \left[\( 1-\cO(\nu^2) \)^{Q} -(Q-1) \cO(\nu^2) \right]^2 \nonumber \\ & \ge 1-\tau' Q\nu^2\;,
\end{align}
for some constant $\tau'=\co(1)$.

Assume now that the state of $\ch_A \otimes \ch_B$ is
$\ket{\psi_0(\beta_f)}=\ket{\phi_0(\beta_f) \, \mathfrak{0}}=\sum_{\sigma=1}^d \sqrt{\pi^\sigma(\beta_f)} \ket{\sigma \, \mathfrak{0}}$. If a
measurement on the $\ket{\sigma}-$basis is performed on $\ch_A$, the
probability of finding the system in an excited state can be
bounded by $d e^{-\beta_f \gamma}$, with $\gamma$ the spectral gap of
$E$. Thus, after the QSA, the total probability
of such an event, which is the error probability for QSA, can be bounded above by
\begin{equation}
\label{probbound}
\cp(\sigma  \not \in \mathbb{S}_0) \le d\, e^{-\beta_f \gamma} + \tau' Q\nu^2 \;,
\end{equation}
as claimed.

\end{appendix}


\begin{thebibliography}{10}
\bibitem{comment1} Note that the rate $\Delta \beta$ can be chosen to
  be $\beta$-dependent if the gap of $M(\beta)$ is used on the right
  side of Eq.~(\ref{rate1}) instead of $\delta$. However, the overall
  implementation complexity of SA will still be dominated by the
  minimum gap.
\end{thebibliography}
\end{document}